\documentclass[A4,11pt]{article}
\usepackage{graphicx}
\usepackage{hyperref}
\usepackage{amssymb, amsmath, amsthm}
\usepackage{colortbl}

\newcommand{\R}{\mathbb R}

\newcommand{\Hh}{{\rm I \hspace{-.8mm} H}}

\def\negr#1{\mbox{\boldmath$#1$}}

\definecolor{lilas}{rgb}{1,0,1}
\definecolor{roxo}{rgb}{0.5,0,1}

\begin{document}

\title
    {Fisher information distance:\\ a geometrical reading\thanks{Partially supported by  FAPESP (Grants 2007/56052-8, 2007/00514-3, 2011/01096-6, 2013/05475-7 and 2013/07375-0), CNPq (Grants 309561/2009-4 and 304032/2010-7) and PRONEX-Optimization.}}

\author
    {S. I. R. Costa%
     \thanks{Institute of Mathematics, University of Campinas, 13083-970, %
	 Campinas, SP, Brazil (sueli@ime.unicamp.br).}\, \and
     S. A. Santos%
     \thanks{Institute of Mathematics, University of Campinas, 13083-970, %
	Campinas, SP, Brazil (sandra@ime.unicamp.br).} \and
    	J. E. Strapasson%
     \thanks{School of Applied Sciences, University of Campinas, 13484-350, %
	Limeira, SP, Brazil (joao.strapasson@fca.unicamp.br).}}

\maketitle


\begin{abstract}
\noindent This paper is a strongly geometrical approach to the Fisher distance, which is a measure of dissimilarity between two probability distribution functions. The Fisher distance, as well as other divergence measures, are also used in many applications to establish a proper data average. The main purpose is to widen the range of possible interpretations and relations of the Fisher distance and its associated geometry for the prospective applications. It focuses on statistical models of the normal probability distribution functions and takes advantage of the connection with the classical hyperbolic geometry to  derive closed forms for the Fisher distance in several cases. Connections with the well-known Kullback-Leibler divergence measure are also devised.  \\ 
\\
\noindent {\bf Keywords}: Fisher distance; information geometry; normal probability distribution functions; Kullback-Leibler divergence; hyperbolic geometry.\\
\\
\noindent {\bf 2010 AMS classification}: 51K99, 94A17, 93B29, 68P20.


\end{abstract}

\newpage

\section{Introduction}
\label{sec:intro}

Information geometry is a research field that has provided framework and  enlarged the perspective of analysis for a wide variety of domains, such as statistical inference, information theory, mathematical programming, neurocomputing, to name a few.  It is an outcome of the investigation of the differential geometric structure on manifolds of probability distributions, with the Riemannian metric defined by the Fisher information matrix  \cite{amari}. Rao's pioneering work  \cite{rao} was subsequently followed by several authors (e.g.  \cite{atkinson,lovric, skovgaard}, among others). We quote \cite{amari} as a general reference for this matter.

Concerning specifically to information theory and signal processing, an important
aspect of the Fisher matrix arises from its trace being related to the
surface area of the typical set associated with a given
probability distribution, whereas the volume of this set is related
to the entropy. This was used to establish connections between
inequalities in information theory and geometric inequalities
(\cite{Max,DemboCover}). 

 The Fisher-Rao metric and the Kullback-Leibler divergence may be used to model experimental data in signal processing. As the underlying Fisher-Rao geometry of Gaussians is hyperbolic without a closed-form equation for the centroids, in \cite[Chap.16]{nielsenBhatia13} the authors have adopted the hyperbolic model centroid approximation, showing its usage in a single-step clustering method. Another recent reference in image processing that also rests upon the hyperbolic geometric structure of the Gaussians is \cite{angulovelasco13}, where morphological operators were formulated for an image model where at each pixel is given a univariate Gaussian distribution, properties of order invariance are explored and an application to morphological
processing of univariate Gaussian distribution-valued images is illustrated.

Current applications of information geometry in statistics include the problem of  dimensionality reduction through information geometric methods on statistical manifolds~\cite{carterTesis} as well as the preparation of samplers for  sequential Monte Carlo techniques \cite{simFilippi}.
In the former, the fact that a manifold of probability density function is often intrinsically lower dimensional than the domain of the data realization provides the background for establishing two methods of dimensionality reduction; the proposed tools are illustrated for case studies on actual patient data sets in the clinical flow cytometric analysis. 
In the latter,  the developed sampler with an information geometric kernel design has attained a higher level of statistical robustness in the inferred parameters of the analyzed dynamical systems than the standard adaptive random walk kernel.

In general, many applications demand a measure of dissimilarity between 
the distributions of the involved objects, or also require the replacement
of a set of data by a proper average or a centroid \cite{hastiebook}. These average representatives could be used, for instance,
as a first step on a classical distance geometry problem \cite{libertietal}.
In these cases, the Fisher distance may apply as well as other dissimilarity measures (\cite{LVAN,nielsenNock09,peterRanga,schutz}).

Our contribution in this paper is to present a geometrical view of the Fisher matrix,
 focusing on the parameters that describe the univariate and the multivariate normal distributions, with the aim of widen the range of possible interpretations for the prospective applications of information geometry in a variety of fields. Our geometrical reading of information geometry fundaments, starting at \S\ref{subsec:hyperb}, allows to employ results from the classical hyperbolic geometry and to derive closed expressions for the Fisher distance in special cases of the multivariate normal distributions. Connections with other dissimilarity measure are also deduced. To enhance the geometric approach, those results are deduced along the text, instead of being displayed in a ``proposition-proof" format. A preliminary summary of some results presented here has appeared in \cite{costa05}.

This text is organized as follows: in Section~\ref{sec:fisheruni} we explore the two dimensional statistical model of the Gaussian (normal) univariate probability distribution  function (PDF). Closed forms for this distance are derived in the most common parameters (cf. (\ref{eq:distsource})-(\ref{eq:distexpectation}) and Figure~\ref{fig:geodesic}) and a relationship with the Kullback-Leibler measure of divergence is presented (see (\ref{eq:KLPQ})-(\ref{eq:distKLPQ}) and Figure~\ref{fig:KB}). Section~\ref{sec:fishermult} is devoted to the Fisher information geometry of the multivariate normal PDF's. For the special cases of the round Gaussian distributions and normal distributions with diagonal covariance matrices, closed forms for the distances are derived (cf. (\ref{eq:distround})  and (\ref{eq:distdiag}), resp.). The Fisher information distance for the general bivariate case is discussed as well (\S\ref{subsec:general}). 

\section{Univariate normal distributions: a geometrical view}
\label{sec:fisheruni}

\subsection{The hyperbolic model of the mean $\times$ standard deviation half-plane}
\label{subsec:hyperb}

The geometric model of the mean $\times$ standard deviation half-plane associates each point in the half upper plane of $\R^2$ with a univariate Gaussian PDF
\[
f(x, \mu, \sigma) = \frac{1}{\sqrt{2 \pi} \sigma} \mbox{ exp} 
\left( \frac{-| x - \mu|^2}{2 \sigma^2} \right).
\]
 Hence, a classic
 parametric space for this family of PDF's is 
\[H = \{ (\mu, \sigma) \in \R^2 \mid \sigma > 0\}.\]

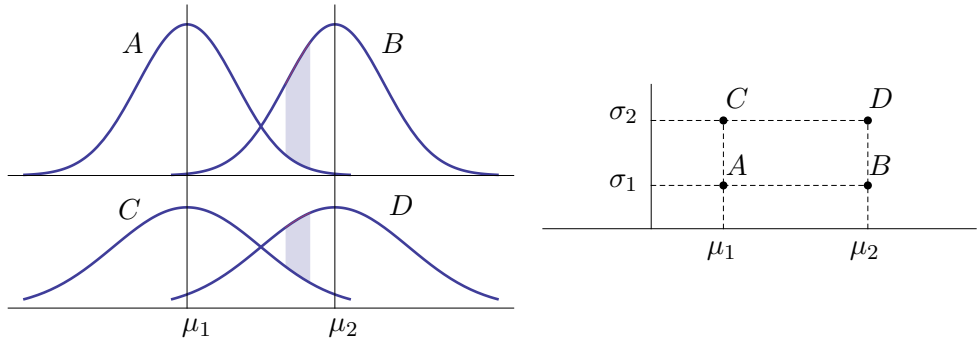
\begin{figure}[ht]
\begin{center}
\parbox{7cm}{
\includegraphics[width=7cm]{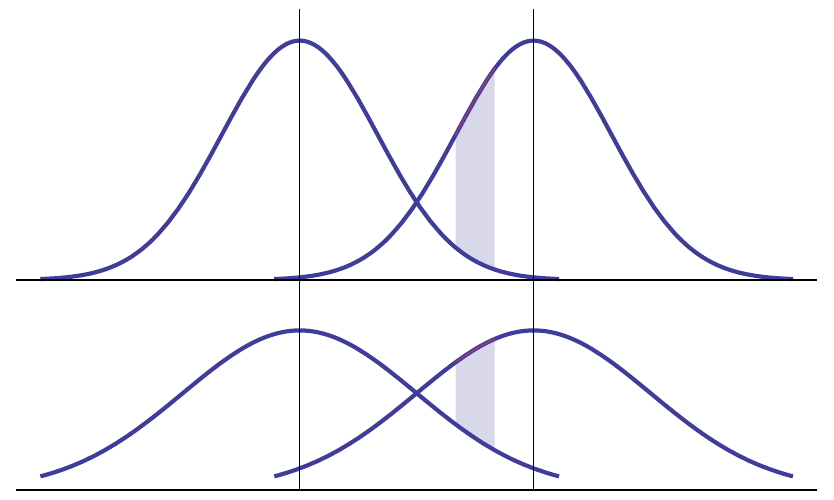} 
\begin{picture}(0,0)(0,0)
\put(2.45,0.3){$\mu_1$}%
\put(4.4,0.3){$\mu_2$}%
\put(1.65,4){$A$}%
\put(5.1,4){$B$}%
\put(1.6,1.8){$C$}%
\put(5.2,1.85){$D$}%
\end{picture}}
\parbox{6cm}{
\includegraphics[width=6cm]{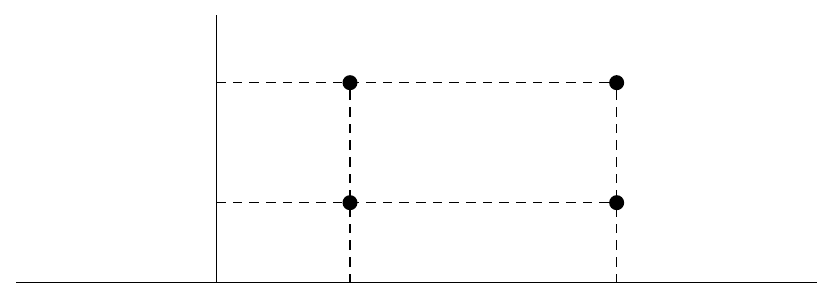} 
\begin{picture}(0,0)(0,0)
\put(2.3,0.3){$\mu_1$}%
\put(4.2,0.3){$\mu_2$}%
\put(1,1.2){$\sigma_1$}%
\put(1,2.1){$\sigma_2$}%
\put(2.55,1.35){$A$}%
\put(4.45,1.35){$B$}%
\put(2.55,2.25){$C$}%
\put(4.45,2.25){$D$}%
\end{picture}}
\parbox{12.5cm}{\caption{Univariate normal distributions and their representations in the $(\mu, \sigma)$ half-plane.}\label{fig:normalUni}}
\end{center}
\end{figure}

A distance between two points $P=(\mu_1,\sigma_1)$ and $Q=(\mu_2,\sigma_2)$ in the half-plane $H$ should reflect the dissimilarity between the associated PDF's.  We will not distinguish the notation of the point $P$ in the parameter space and its associated PDF $f(x,P)$.

A comparison between univariate normal distributions is illustrated in  Figure~\ref{fig:normalUni}. By fixing the means and increasing the standard deviation, we can see that the dissimilarity between the probabilities attached to the same interval concerning the PDF's associated with $C$ and $D$ is smaller than the one between the PDF's associated with $A$ and $B$~(left). This means that the distance between points in the upper half-plane~(right) representing normal distributions cannot be Euclidean. Moreover, we can observe that such a metric must vary with the inverse of the standard deviation $\sigma$. The points $C$ and $D$ should be closer to each other than the points $A$ and $B$, reflecting that the pair of distributions $A$ and $B$ is more dissimilar than the pair $C$ and $D$. 

A proper distance arises from the Fisher information matrix, which
is a measure of the amount of information of the location parameter
(\cite{CoverThomas}, ch. 12). For univariate distributions parametrized 
by an $n$-dimensional space, 
the coefficients of this matrix, which define a metric, are
calculated as the expectation of a product involving partial
derivatives of the logarithm
of the PDF's:%

\[
g_{ij}(\negr{\beta})=\int\limits_{-\infty}^{\infty}f(x,\negr{\beta})\frac{\partial\ln
f(x,\negr{\beta})}{\partial\beta_{i}}\frac{\partial\ln
f(x,\negr{\beta})}{\partial \beta_{j}}dx.
\]

A metric matrix $G=(g_{ij})$ defines an inner product as follows:
\[
\langle u, v \rangle_G = u^T (g_{ij}) v \quad \mbox{and} \quad \|u\|_G = \sqrt{ \langle u, u \rangle_G}.
\]

The distance between two points $P, Q$ is given by the number which is the minimum of the lengths
of all the piecewise smooth paths $\gamma_P^Q$ joining these two points.The length of a path $\gamma(t)$ is calculated by using the inner 
product $\langle \cdot , \cdot \rangle_G$:
\[
\mbox{Length of } \gamma = \int_\gamma ds = \int_\gamma \| \gamma^{\prime}(t)\|_G dt 
\]
and so 
\[
d_G(P,Q)= \min_{\gamma_P^Q} \{ \mbox{Length of }  \gamma\}.
\]

A curve that encompasses this shortest path is a \emph{geodesic}.

In the univariate normally distributed case described above we
have $\negr{\beta}=(\beta_{1},\beta_{2})=\left(  \mu,\sigma\right)  $
and it can be easily deduced that the Fisher information matrix is%
\begin{equation}
\left[  g_{ij}(\mu, \sigma)\right]_F  =\left[
\begin{array}
[c]{cc}%
\frac{1}{\sigma^{2}} & 0\\
0 & \frac{2}{\sigma^{2}}%
\end{array}
\right]
\label{eq:fishermatrix}
\end{equation}
so that the expression for the metric is
\begin{equation}
ds^2_F = \frac{d\mu^2+ 2 d\sigma^2}{\sigma^2}.
\label{eq:metricstar}
\end{equation}

The Fisher distance is the one associated with the Fisher information matrix (\ref{eq:fishermatrix}). 
In order to express such a notion of distance and to characterize the geometry in the plane $\Hh^2_F$,
we analyze its analogies with the well-known Poincar\'e half-plane $\Hh^2$, a model for the hyperbolic geometry, the metric of which is given by the matrix
\begin{equation}
\left[  g_{ij}\right]_H  =\left[
\begin{array}
[c]{cc}%
\frac{1}{\sigma^{2}} & 0\\
0 & \frac{1}{\sigma^{2}}%
\end{array}
\right].
\label{eq:poincarematrix}
\end{equation}

The  inner product associated with the Fisher matrix (\ref{eq:fishermatrix}) will be denoted by $\langle \cdot,\cdot\rangle_F$ and the distance between $P=(\mu_{1},\sigma_{1})$ and
$Q=(\mu_{2},\sigma_{2})$ in the upper half-plane $\Hh^2_F$, by $d_F(P,Q)$. The distance in the Poincar\'e half-plane induced by (\ref{eq:poincarematrix}) will be denoted by $d_H(P,Q)$. 
By considering the similarity mapping $\Psi: \Hh^2_F \rightarrow \Hh^2$ defined by $\Psi(\mu, \sigma) = (\mu/\sqrt{2},\sigma)$, we can see that
\begin{equation}
d_{F}((\mu_{1},\sigma_{1}),(\mu_{2},\sigma_{2}))=\sqrt{2}d_{H}\left(\left(\frac{\mu_{1}%
}{\sqrt{2}},\sigma_{1}\right),\left(\frac{\mu_{2}}{\sqrt{2}},\sigma_{2}\right)\right),
\label{eq:distFH}
\end{equation}

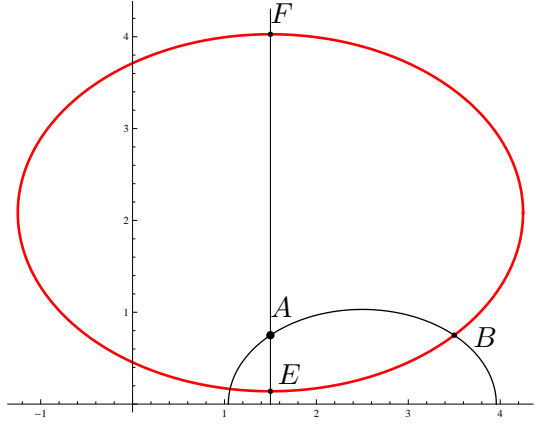
\begin{figure}[ht]
\begin{center}
\parbox{7cm}{
\includegraphics[width=7cm]{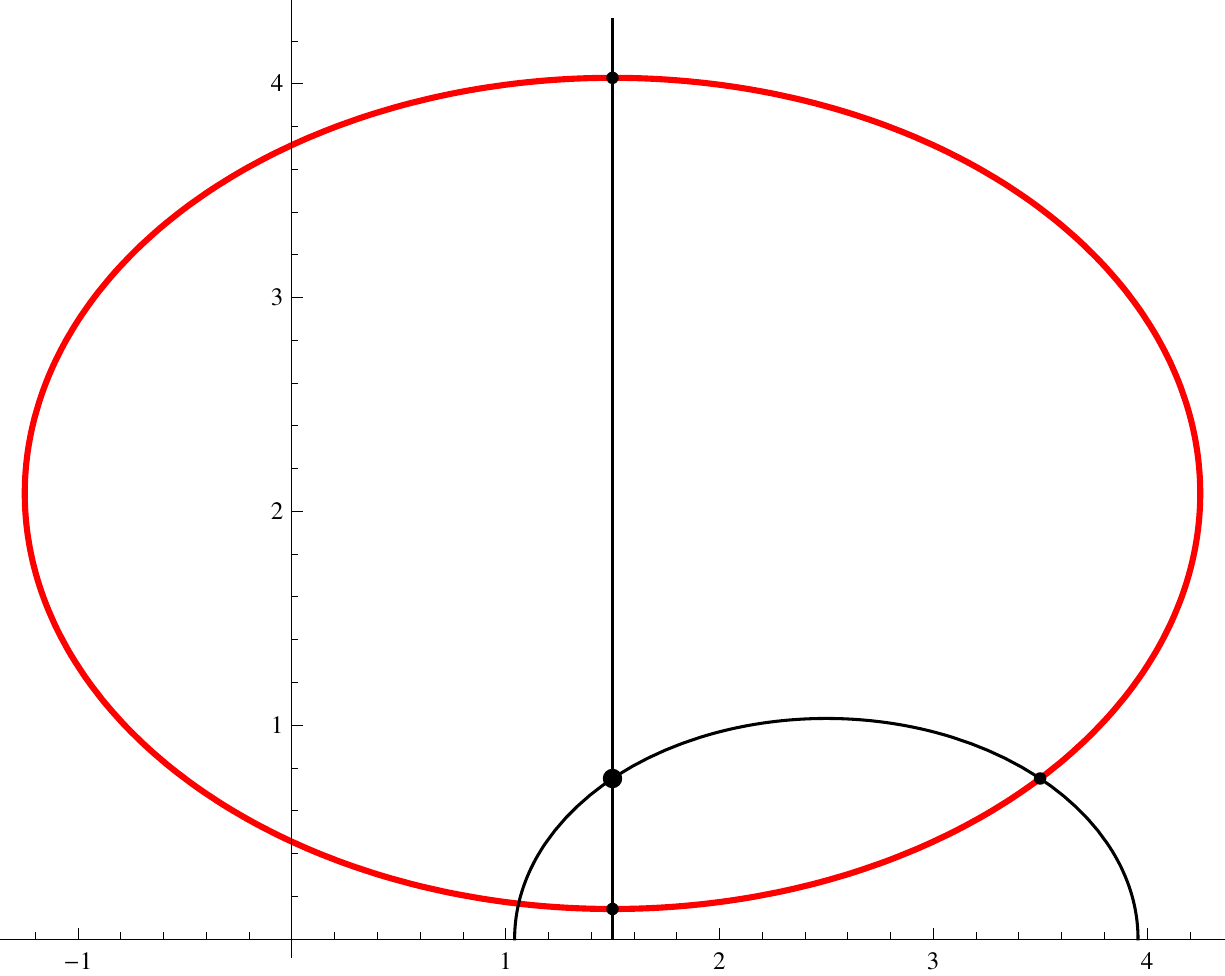} 
\begin{picture}(0,0)(0,0)
\put(3.5,1.9){$A$}%
\put(3.5,5.8){$F$}%
\put(3.6,1){$E$}%
\put(6.2,1.5){$B$}%
\end{picture}}
\parbox{12.5cm}{\caption{A Fisher circle centered at $A$ and geodesic arcs $AB$, $AF$ and $AE$, with $d_F(A,B)=d_F(A,F)=d_F(A,E)$.}\label{fig:circFisher}}
\end{center}
\end{figure}

Besides, the geodesics in $\Hh^2_F$ are the inverse image, by $\Psi$, of the geodesics in $ \Hh^2$.
Vertical half-lines and half-circles centered at $\sigma=0$ are the geodesics in 
 $\Hh^2$ (see, eg. \cite[Ch.7]{beardon}). Hence, the geodesics in $ \Hh^2_F$ are half-lines and half-ellipses centered at $\sigma=0$, with eccentricity $1/\sqrt{2}$. We can also assert that a circle in the Fisher distance is an ellipse with the same eccentricity and its center is below the Euclidean center. Figure~\ref{fig:circFisher} shows the Fisher circle centered at $A=(1.5,0.75)$ and radius 2.3769, and the geodesics connecting the center to points $B$, $E$ and $F$ on the circle.

\begin{figure}[ht]
\begin{center}
\parbox{3.5cm}{
\includegraphics[height=3.7cm]{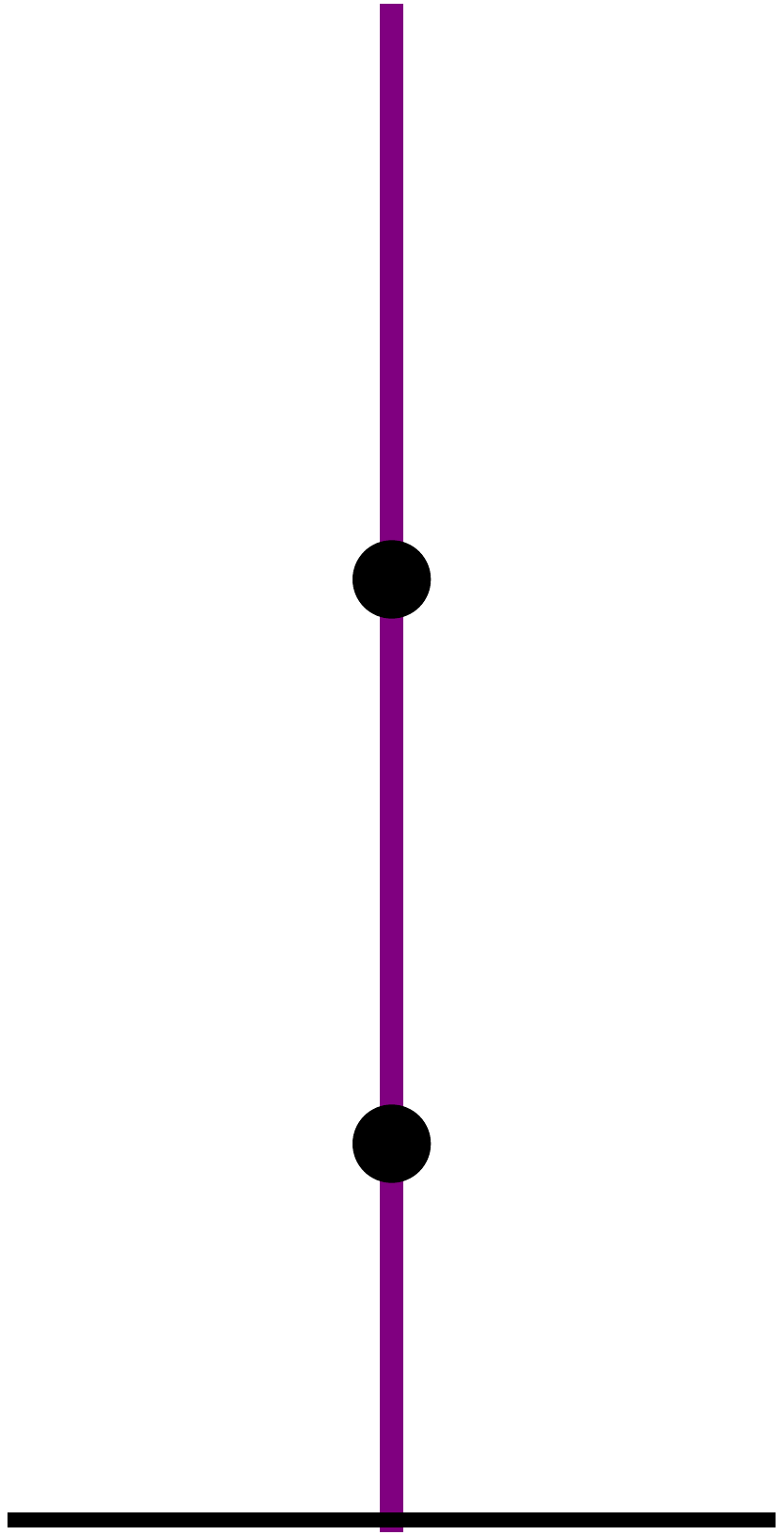} \\[20pt]
\begin{picture}(0,0)(0,0)
\put(0.9,0.8){$P_\infty$}%
\put(1.1,2){$P$}%
\put(1.1,3.5){$Q$}%
\put(0.9,5.1){$Q_\infty$}%
\end{picture}}
\parbox{7.5cm}{
\includegraphics[width=7.5cm]{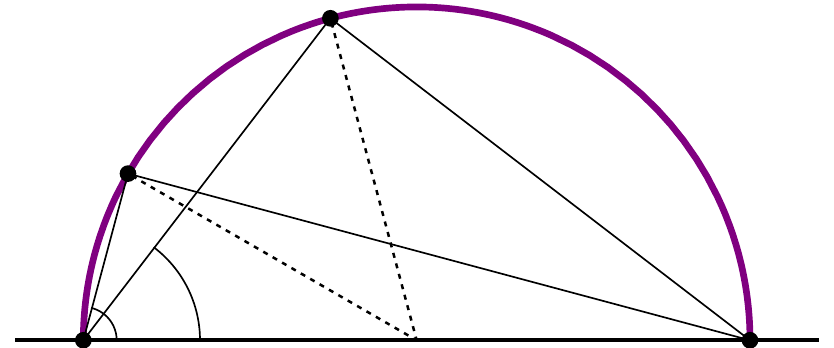} 
\begin{picture}(0,0)(0,0)
\put(0.6,0.15){$P_\infty$}%
\put(0.7,2.1){$P$}%
\put(2.55,3.6){$Q$}%
\put(6.5,0.15){$Q_\infty$}%
\put(1.1,0.8){$\frac{\alpha_P}{2}$}%
\put(1.7,1.05){$\frac{\alpha_Q}{2}$}%
\end{picture}}
\parbox{12cm}{\caption{Elements to compute the distance $d_H(P,Q)$, in case the points $P, Q \in \Hh^2$ are vertically aligned (left) or not (right).}\label{fig:crossratio}}
\end{center}
\end{figure}
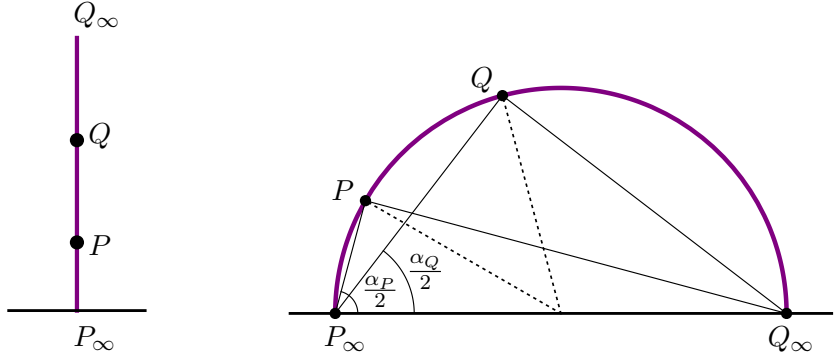

The distance between two points in the Poincar\'e half-plane can be expressed by the logarithm of the cross-ratio between these two points 
and the points at the infinite:
\[
d_H(P,Q) = \ln(P_\infty,P,Q,Q_\infty).
\]
It can be stated by the following formulas, considering $P$ and $Q$ as vertical lined or not,
as illustrated in Figure~\ref{fig:crossratio}, respectively:
\[
d_H(P,Q) =\ln \left( \frac{\sigma_Q}{\sigma_P}\right) \;\mbox{or } \;
d_H(P,Q) = \ln \left( \frac{PQ_\infty}{PP_\infty} \cdot\frac{QP_\infty}{QQ_\infty}\right) = 
\ln\left( \frac{\tan\left(\frac{\alpha_P}{2}\right)}{\tan\left(\frac{\alpha_Q}{2}\right)}\right).
\]

By recalling that the Fisher distance $d_{F}$ and the hyperbolic distance $d_{H}$ are related
by (\ref{eq:distFH}) we obtain the following closed expression for the Fisher
information distance:
\begin{eqnarray}
d_{F}((\mu_{1},\sigma_{1}),(\mu_{2},\sigma_{2}))  = 
 \sqrt{2}\ln\dfrac{\left\vert
(\frac{\mu_{1}}{\sqrt{2}},\sigma_{1})-(\frac{\mu_{2}}{\sqrt{2}},-\sigma
_{2})\right\vert +\left\vert
(\frac{\mu_{1}}{\sqrt{2}},\sigma_{1})-(\frac
{\mu_{2}}{\sqrt{2}},\sigma_{2})\right\vert }{\left\vert (\frac{\mu_{1}}%
{\sqrt{2}},\sigma_{1})-(\frac{\mu_{2}}{\sqrt{2}},-\sigma_{2})\right\vert
-\left\vert (\frac{\mu_{1}}{\sqrt{2}},\sigma_{1})-(\frac{\mu_{2}}{\sqrt{2}%
},\sigma_{2})\right\vert }%
\label{eq:distfisher}\\[10pt]
 =  \sqrt{2}\ln \left(
\dfrac{{\cal F}((\mu_{1},\sigma_{1}),(\mu_{2},\sigma_{2})) + (\mu_1-\mu_2)^2 + 2(\sigma_1^2+ \sigma_2^2)}{4 \sigma_1 \sigma_2}
\right)
\label{eq:distfisher2}
\end{eqnarray}
where
\[
{\cal F}((\mu_{1},\sigma_{1}),(\mu_{2},\sigma_{2}))=
\sqrt{((\mu_1-\mu_2)^2 +2 (\sigma_1- \sigma_2)^2)((\mu_1-\mu_2)^2 +2 (\sigma_1+ \sigma_2)^2)}.
\]

\begin{figure}[ht]
\begin{center}
\parbox{7cm}{
\includegraphics[width=7cm]{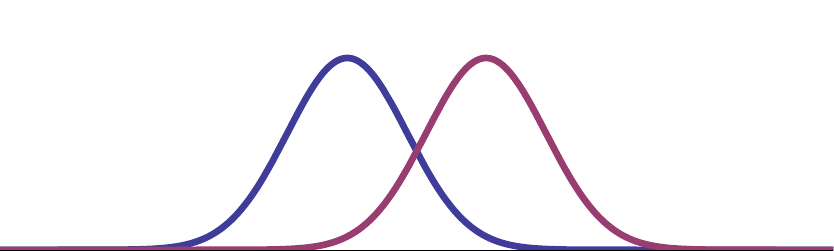} 
\includegraphics[width=7cm]{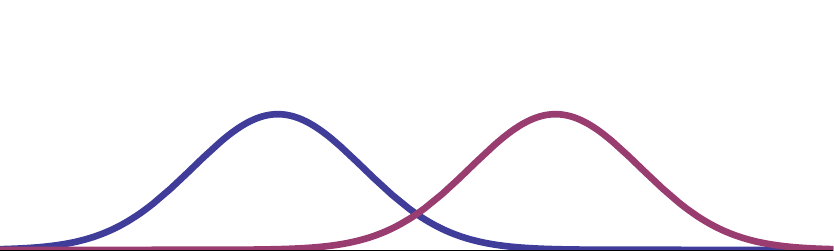} 
\begin{picture}(0,0)(0,0)
\put(2.1,4){$A$}%
\put(4.6,4){$B$}%
\put(1.4,1.4){$C$}%
\put(5.4,1.4){$D$}%
\end{picture}}\hspace{1cm}
\parbox{5cm}{
\includegraphics[width=5cm]{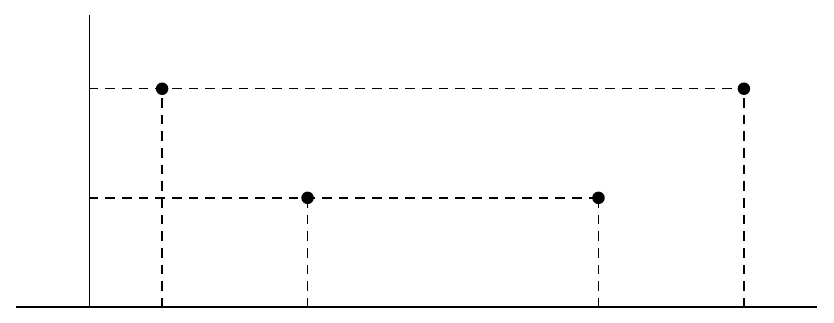} 
\begin{picture}(0,0)(0,0)
\put(4.5,0.35){$\mu$}%
\put(0.2,2.3){$\sigma$}%
\put(1.7,1.4){$A$}%
\put(3.5,1.4){$B$}%
\put(0.8,2.1){$C$}%
\put(4.4,2.1){$D$}%
\end{picture}}
\parbox{12.5cm}{\caption{Equidistant pairs in Fisher metric: $d_H(A,B) = d_F(C,D)=2.37687$, where $A=(1.5,0.75)$, $B=(3.5,0.75)$ and $C=(0.5,1.5)$, $D=(4.5,1.5)$.}\label{fig:equidist}}
\end{center}
\end{figure}
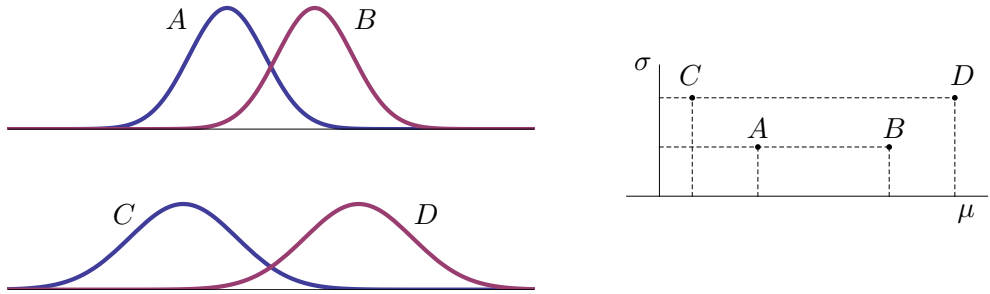

Figure \ref{fig:equidist} illustrates two distinct pairs of Gaussian distributions which are equidistant with the Fisher metric. Moreover, from the relations (\ref{eq:distfisher})-(\ref{eq:distfisher2}) we can deduce facts of the geometry of  the upper half plane with the Fisher metric: it is hyperbolic with
constant curvature equal to $-\frac{1}{2}$ and the shortest path
between the representatives of two normal distributions is either
on a vertical line or on a half ellipse (see Figure~\ref{fig:geodesic}(a)). 

 The Fisher distance between two PDF's $P=(\mu, \sigma_1)$ and  $Q=(\mu, \sigma_2)$  is
\begin{equation}
 d_F(P,Q)=\sqrt{2} |\ln(\sigma_2/\sigma_1)|
\label{eq:verticaldistance}
\end{equation}
 and the vertical line connecting $P$ and $Q$ is a geodesic in the Fisher half-plane. On the other hand, the geodesic connecting $P=(\mu_1, \sigma)$ and  $Q=(\mu_2, \sigma)$ associated with two normal PDF's with the same variance is not the horizontal line connecting these points (the shortest path is contained in a half-ellipse). Indeed,  
\begin{equation}
d_F(P,Q) = \sqrt{2}\ln\left(
\frac{4 \sigma^2 + (\mu_1-\mu_2)^2+|\mu_1-\mu_2|\sqrt{8 \sigma^2 + (\mu_1-\mu_2)^2 }}{4 \sigma^2}
\right) < \frac{|\mu_2-\mu_1|}{\sigma}.
\label{eq:horizontal}
\end{equation}
The expression on the right of (\ref{eq:horizontal}) is the length of the horizontal segment joining $P$ and $Q$. Nevertheless, in case just normal PDF's with constant variance are considered, the expression on the right of (\ref{eq:horizontal}) is a proper distance.

\begin{figure}[ht]
\begin{center}
\parbox{6cm}{
\includegraphics[width=6cm]{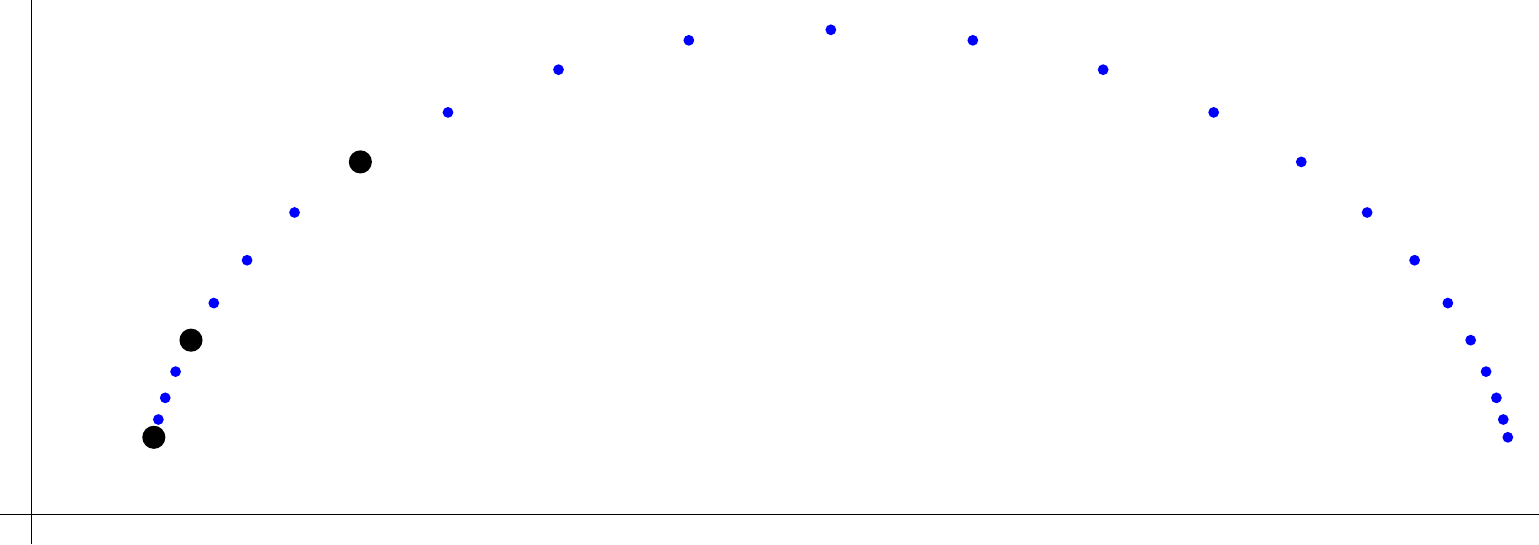} 
\begin{picture}(0,0)(0,0)
\put(1,2){$A$}%
\put(0.2,1){$Q$}%
\put(0.3,1.5){$M$}%
\end{picture}}
\parbox{12.5cm}{\caption{The Fisher average between distributions $A=(1.5, .75)$ and $Q=(1.0610, 0.1646)$  is  $M=  (1.1400, 0.3711)$. The plotted points form a polygonal with equal Fisher length segments.}\label{fig:average}}
\end{center}
\end{figure}
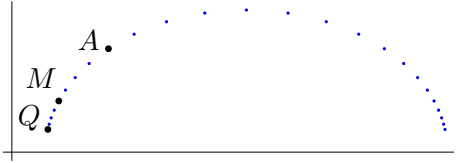

It is worth mentioning that the Fisher metric can also be used to establish the concept of \emph{average} distribution between two given 
distributions $A$ and $Q$. This is determined by the point $M$ on the geodesic segment joining $A$ and $Q$ and which is equidistant to these points in Figure~\ref{fig:average}.

\subsection{Univariate normal distributions described in other usual parameters}
\label{subsec:otherparam}

Univariate normal distributions may be also described by means of the so-called \emph{source} $(\lambda_1, \lambda_2) \in \R \times \R_+$, \emph{natural} $(\theta_1, \theta_2) \in \R \times \R_-$ and \emph{expectation} parameters $(\eta_1, \eta_2)  \in \R \times \R_+$, respectively defined  by
\[
(\lambda_1, \lambda_2) = (\mu, \sigma^2),
\]
\[
(\theta_1, \theta_2)  = \left(\frac{\mu}{\sigma^2}, \frac{-1}{2 \sigma^2}  \right) 
\]
and
\[
(\eta_1, \eta_2)  = \left(\mu,  \sigma^2+ \mu^2  \right).
\]
Therefore, 
\[
(\mu, \sigma) = (\lambda_1, \sqrt{\lambda_2}) =  \left(\frac{- \theta_1}{2 \theta_2}, \frac{1}{\sqrt{-2 \theta_2}}  \right)  = \left(\eta_1, \sqrt{\eta_2 - \eta_1^2}  \right)
\]
and expressions  (\ref{eq:distfisher})-(\ref{eq:distfisher2})  may be restated, for the source parameters, as
\begin{equation}
\begin{array}{l}
d_{F}((\lambda_{11}, \sqrt{\lambda_{21}}),(\lambda_{12}, \sqrt{\lambda_{22}}))  = d_{\lambda}((\lambda_{11},\lambda_{21}),(\lambda_{12},\lambda_{22}))  = \\[5pt]
 \sqrt{2}\ln \left(-
\dfrac{
\sqrt{(\lambda_{11}-\lambda_{12})^2 + 2(\sqrt{\lambda_{21}}-\sqrt{\lambda_{22}})^2}+
\sqrt{(\lambda_{11}-\lambda_{12})^2 + 2(\sqrt{\lambda_{21}}+\sqrt{\lambda_{22}})^2}}
{\sqrt{(\lambda_{11}-\lambda_{12})^2 + 2(\sqrt{\lambda_{21}}-\sqrt{\lambda_{22}})^2}-
\sqrt{(\lambda_{11}-\lambda_{12})^2 + 2(\sqrt{\lambda_{21}}+\sqrt{\lambda_{22}})^2}}
\right),
\end{array}
\label{eq:distsource}
\end{equation}
for the natural parameters as
\begin{equation}
\begin{array}{l}
d_{F}\left(\left(\dfrac{- \theta_{11}}{2 \theta_{21}}, \dfrac{1}{\sqrt{-2 \theta_{21}}}\right),
\left(\dfrac{- \theta_{12}}{2 \theta_{22}}, \dfrac{1}{\sqrt{-2 \theta_{22}}}\right)\right)  = d_{\theta}((\theta_{11},\theta_{21}),(\theta_{12},\theta_{22}))  = \\[15pt]
 \sqrt{2}\ln \left(-
\dfrac{
\sqrt{4 \left(\frac{1}{\sqrt{-\theta_{21}}}-\frac{1}{\sqrt{-\theta_{21}}} \right)^2 +
\left(\frac{\theta_{11}}{\theta_{21}} -\frac{\theta_{12}}{\theta_{22}} \right)^2}+
\sqrt{4 \left(\frac{1}{\sqrt{-\theta_{22}}}+\frac{1}{\sqrt{-\theta_{21}}} \right)^2 +
\left(\frac{\theta_{11}}{\theta_{21}} -\frac{\theta_{12}}{\theta_{22}} \right)^2}}
{\sqrt{4 \left(\frac{1}{\sqrt{-\theta_{21}}}-\frac{1}{\sqrt{-\theta_{21}}} \right)^2 +
\left(\frac{\theta_{11}}{\theta_{21}} -\frac{\theta_{12}}{\theta_{22}} \right)^2}-
\sqrt{4 \left(\frac{1}{\sqrt{-\theta_{22}}}+\frac{1}{\sqrt{-\theta_{21}}} \right)^2 +
\left(\frac{\theta_{11}}{\theta_{21}} -\frac{\theta_{12}}{\theta_{22}} \right)^2}}
\right)
\end{array}
\label{eq:distnatural}
\end{equation}
and for the expectation parameters as
{\footnotesize \begin{equation}
\begin{array}{l}
d_{F}(( \eta_{11}, \sqrt{\eta_{21}- \eta_{11}^2}),
( \eta_{12}, \sqrt{\eta_{22}- \eta_{12}^2}))=
  d_{\eta}((\eta_{11},\eta_{21}),(\eta_{12},\eta_{22}))  = \\[5pt]
 \sqrt{2}\ln \left(-
\dfrac{
\sqrt{(\eta_{11}-\eta_{12})^2 + 2\left( \sqrt{\eta_{21}- \eta_{11}^2} - \sqrt{\eta_{22}- \eta_{12}^2}\right)^2}+
\sqrt{(\eta_{11}-\eta_{12})^2 + 2\left( \sqrt{\eta_{21}- \eta_{11}^2} + \sqrt{\eta_{22}- \eta_{12}^2}\right)^2}}
{\sqrt{(\eta_{11}-\eta_{12})^2 + 2\left( \sqrt{\eta_{21}- \eta_{11}^2} - \sqrt{\eta_{22}- \eta_{12}^2}\right)^2}-
\sqrt{(\eta_{11}-\eta_{12})^2 + 2\left( \sqrt{\eta_{21}- \eta_{11}^2} + \sqrt{\eta_{22}- \eta_{12}^2}\right)^2}}
\right).
\end{array}
\label{eq:distexpectation}
\end{equation}}

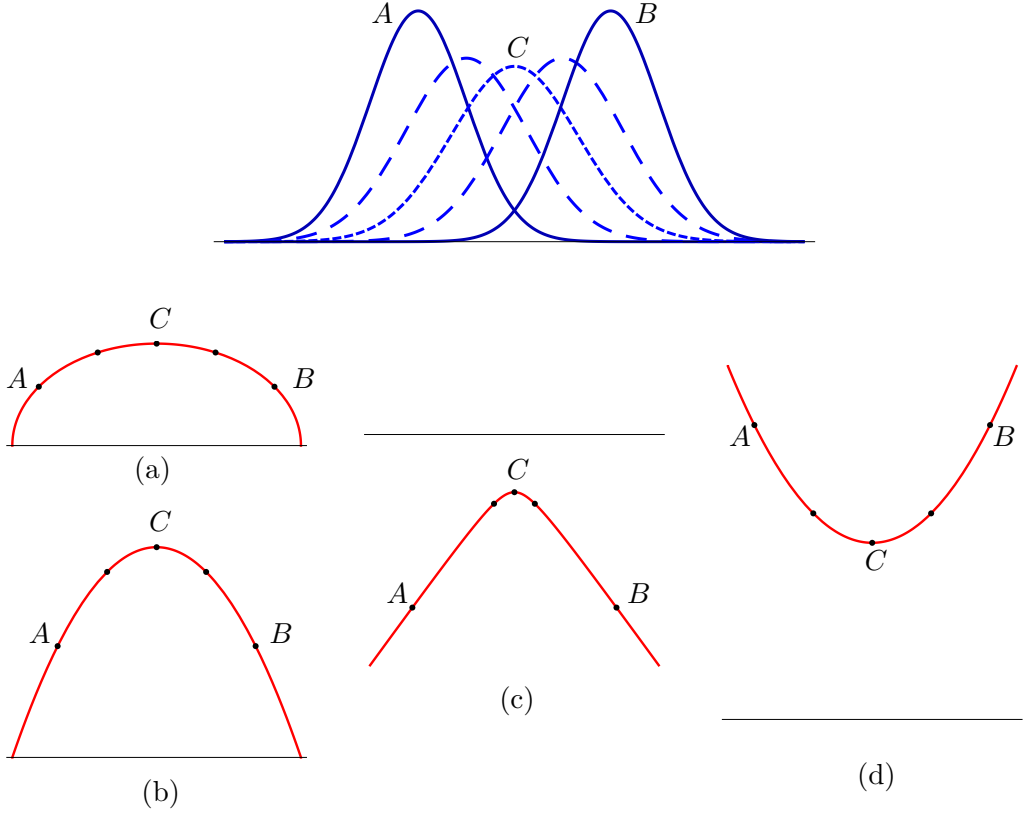
\begin{figure}[ht]
\begin{center}
\parbox{8cm} {
\includegraphics[width=8cm]{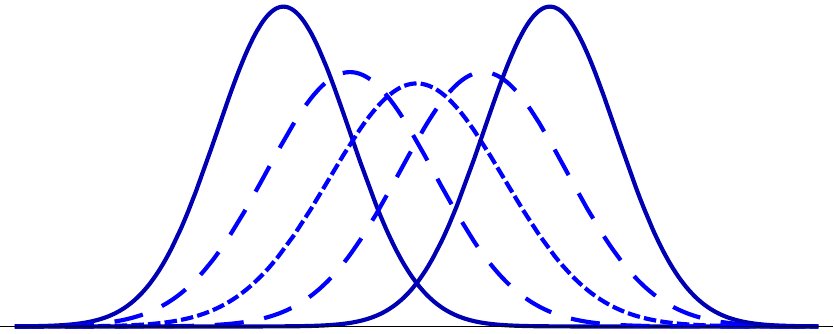}
\begin{picture}(0,0)(0,0)
\put(2.1,3.5){$A$}%
\put(3.9,3.05){$C$}%
\put(5.6,3.5){$B$}%
\end{picture}}
\\[20pt]
\parbox{4cm} {
\includegraphics[width=4cm]{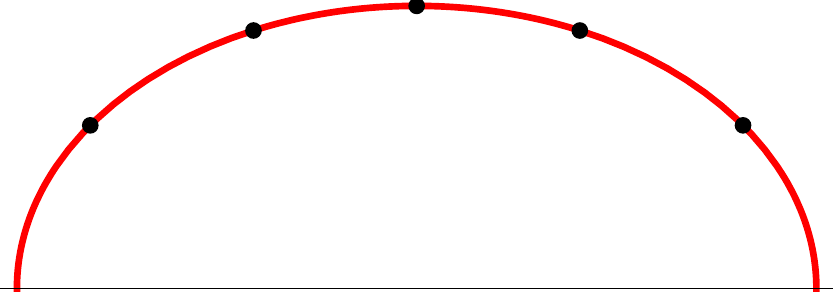} \\[35pt]
\includegraphics[width=4cm]{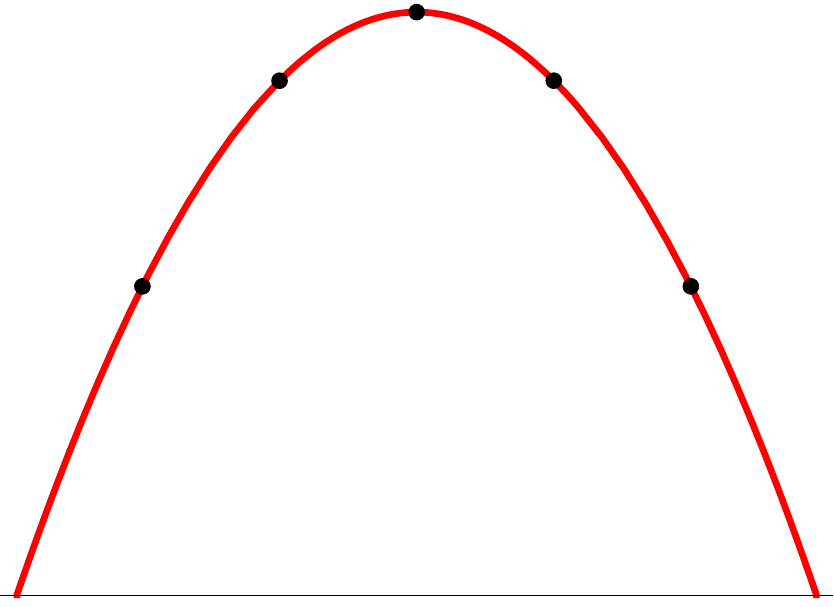}
\begin{picture}(0,0)(0,0)
\put(1.7,4.3){(a)}%
\put(0,5.5){$A$}%
\put(1.9,6.3){$C$}%
\put(3.8,5.5){$B$}%
\put(1.8,0){(b)}%
\put(0.3,2.1){$A$}%
\put(1.9,3.6){$C$}%
\put(3.5,2.1){$B$}%
\end{picture}}
\hspace{0.5cm}
\parbox{4cm} {
\includegraphics[width=4cm]{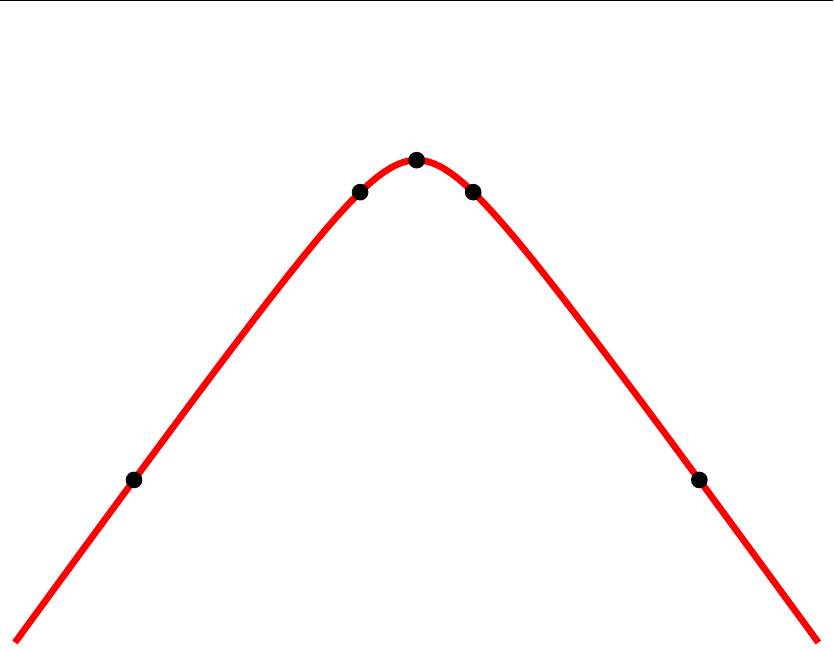} 
\begin{picture}(0,0)(0,0)
\put(1.8,0){(c)}%
\put(0.3,1.4){$A$}%
\put(1.9,3.05){$C$}%
\put(3.5,1.4){$B$}%
\end{picture}}
\hspace{0.5cm}
\parbox{4cm} {
\includegraphics[width=4cm]{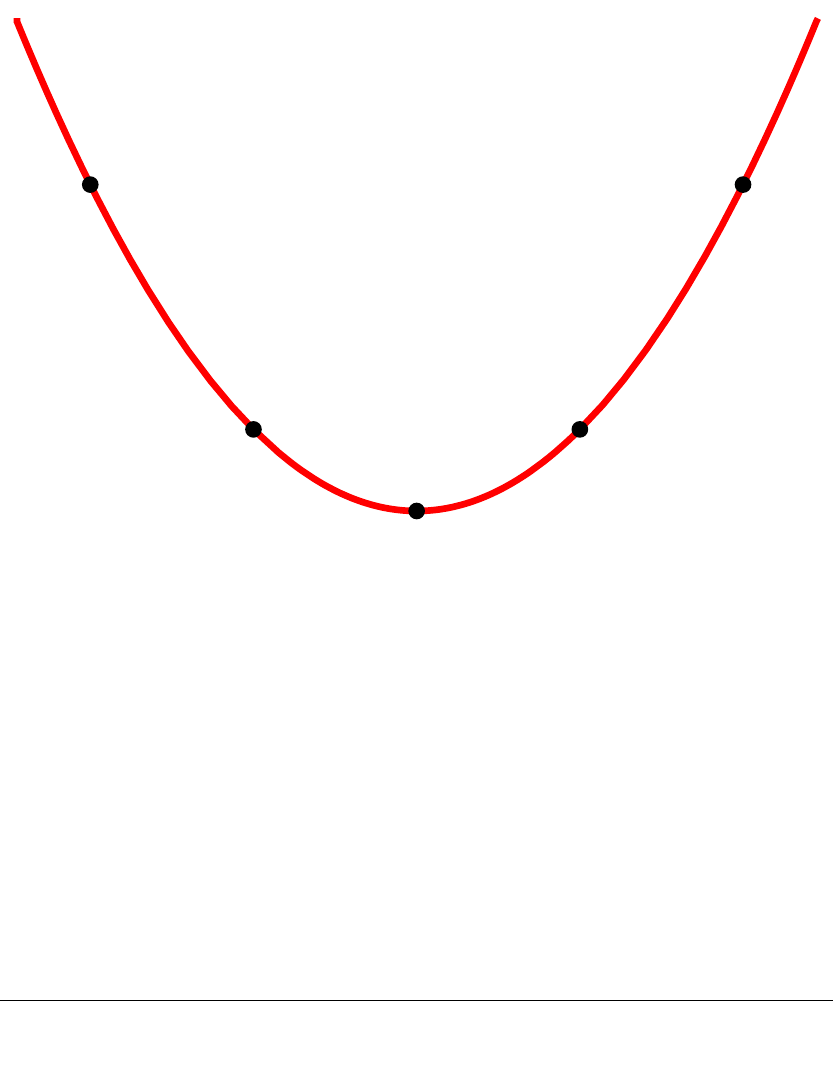}
\begin{picture}(0,0)(0,0)
\put(1.8,0){(d)}%
\put(0.1,4.5){$A$}%
\put(1.9,2.85){$C$}%
\put(3.6,4.5){$B$}%
\end{picture}}
{\caption{Shortest path between the normal distributions $A$ and $B$ in the distinct half-planes: (a) Classic parameters $(\mu,\sigma)$ -- mean $\times$ standard deviation;
 (b) Source parameters $(\mu,\sigma^2)$ -- mean $\times$ variance;
(c) Natural parameters $(\theta_1,\theta_2)=(\frac{\mu}{\sigma^2}, \frac{-1}{2\sigma^2})$ and
(d) Expectation parameters $(\eta_1,\eta_2)=(\mu, \mu^2+\sigma^2)$. 
}\label{fig:geodesic}}
\end{center}
\end{figure}

The shortest path between two normal distributions is depicted in Figure~\ref{fig:geodesic} for the four distinct half-planes, described by the classic~(a), the source (b), the natural~(c) and the expectation parameters~(d).
Besides the half-ellipse that contains the shortest path in the classic mean $\times$ standard deviation half-plane,  the shortest path in the source mean $\times$ variance  and in the expectation half-planes  are described by arc of parabolas, whereas an arc of a half-hyperbola contains the shortest path in the natural half-plane.

\subsection{The Kullback-Leibler divergence and the Fisher distance}
\label{subsec:KL}

Another measure of dissimilarity between two PDF's is the 
Kullback-Leibler divergence \cite{KL}, which is  used in information theory and commonly referred to as the relative entropy of a probability distribution. It is not a distance neither a symmetric measure. In what follows  we discuss its relation with the Fisher distance in the case of univariate normal distributions. Its expression in this case is:
\[
KL((\mu_1,\sigma_1)||(\mu_2,\sigma_2)) =\frac{1}{2} \left(2 \ln \left[\frac{\sigma _2}{\sigma _1}\right]+\frac{\sigma _1^2}{\sigma _2^2}+\frac{\left(\mu _1-\mu _2\right){}^2}{\sigma _2^2} -1\right)
\]

A symmetrized version of this measure,  
\begin{equation}
\begin{array}{rcl}
d_{KL}((\mu_1,\sigma_1),(\mu_2,\sigma_2)) &=&\sqrt{KL((\mu_1,\sigma_1)||(\mu_2,\sigma_2)) +KL((\mu_2,\sigma_2)||(\mu_1,\sigma_1)) } \\[10pt]
&=&\sqrt{\frac{1}{2} \left(-2+\frac{\left(\mu _1-\mu _2\right){}^2}{\sigma _1^2}+\frac{\sigma _1^2}{\sigma _2^2}+\frac{\left(\mu _1-\mu _2\right){}^2}{\sigma _2^2}+\frac{\sigma _2^2}{\sigma _1^2}\right)},
\end{array}
\label{eq:distKL}
\end{equation}
is also used.

If the points in the parameter space are vertically aligned ($P=(\mu,\sigma_1)$ and $Q=(\mu,\sigma_2)$), the Fisher distance is $d=d_F(P,Q)=\sqrt{2} \ln(\frac{\sigma_2}{\sigma_1})$, from what we get an expression of the Kullback-Leibler divergences in terms of $d$:
\begin{equation}
KL(P||Q)=g(d)=\frac{1}{2} \left(e^{-\sqrt{2} d}+2 \ln \left(e^{\frac{d}{\sqrt{2}}}\right)-1\right), \quad KL(Q||P)=g(-d)
\label{eq:KLPQ}
\end{equation}
and
\begin{equation}
 d_{KL}(P,Q)=\sqrt{\dfrac{e^{\sqrt{2} d} +e^{-\sqrt{2} d}}{2} -1}=\sqrt{\cosh(\sqrt{2} d)-1}. 
\label{eq:distKLPQ}
\end{equation}

\begin{figure}[ht]
\begin{center}
\includegraphics[width=5cm]{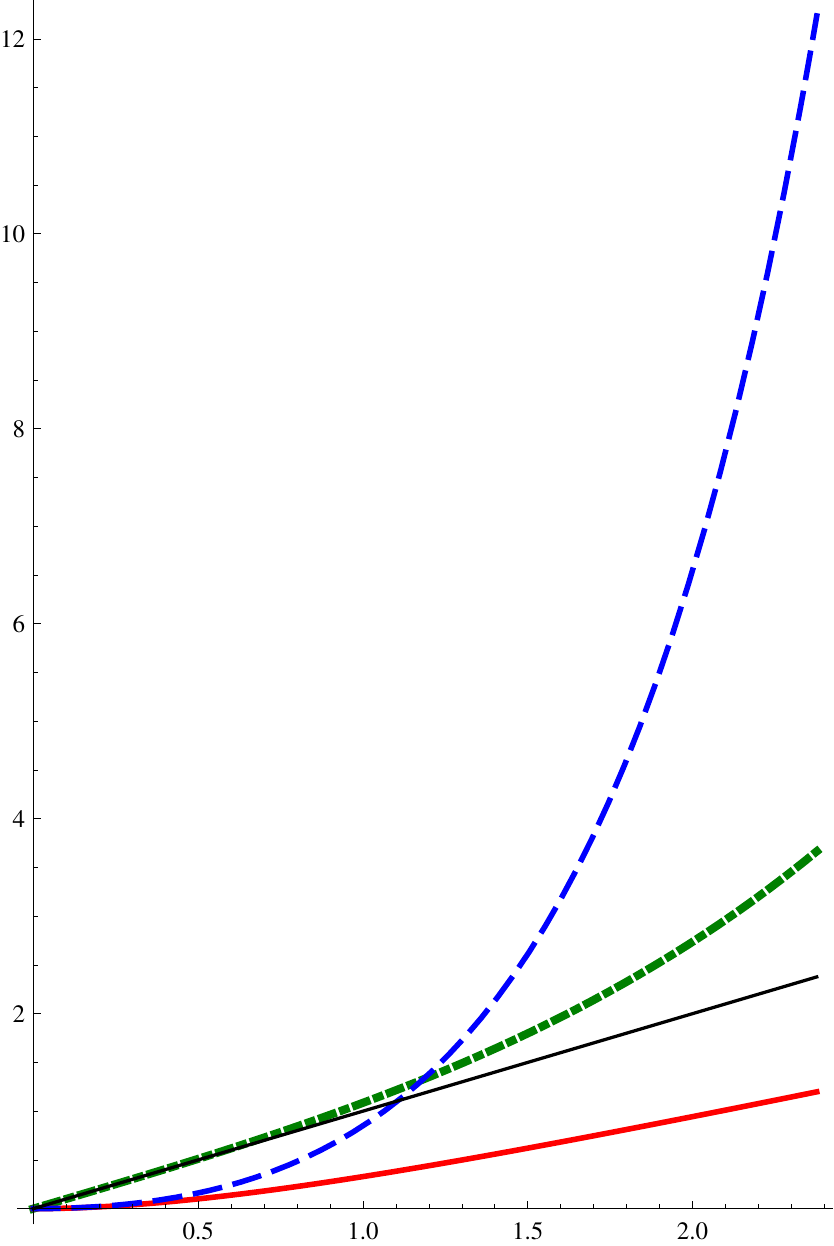} \hspace{1cm}
\includegraphics[width=5cm]{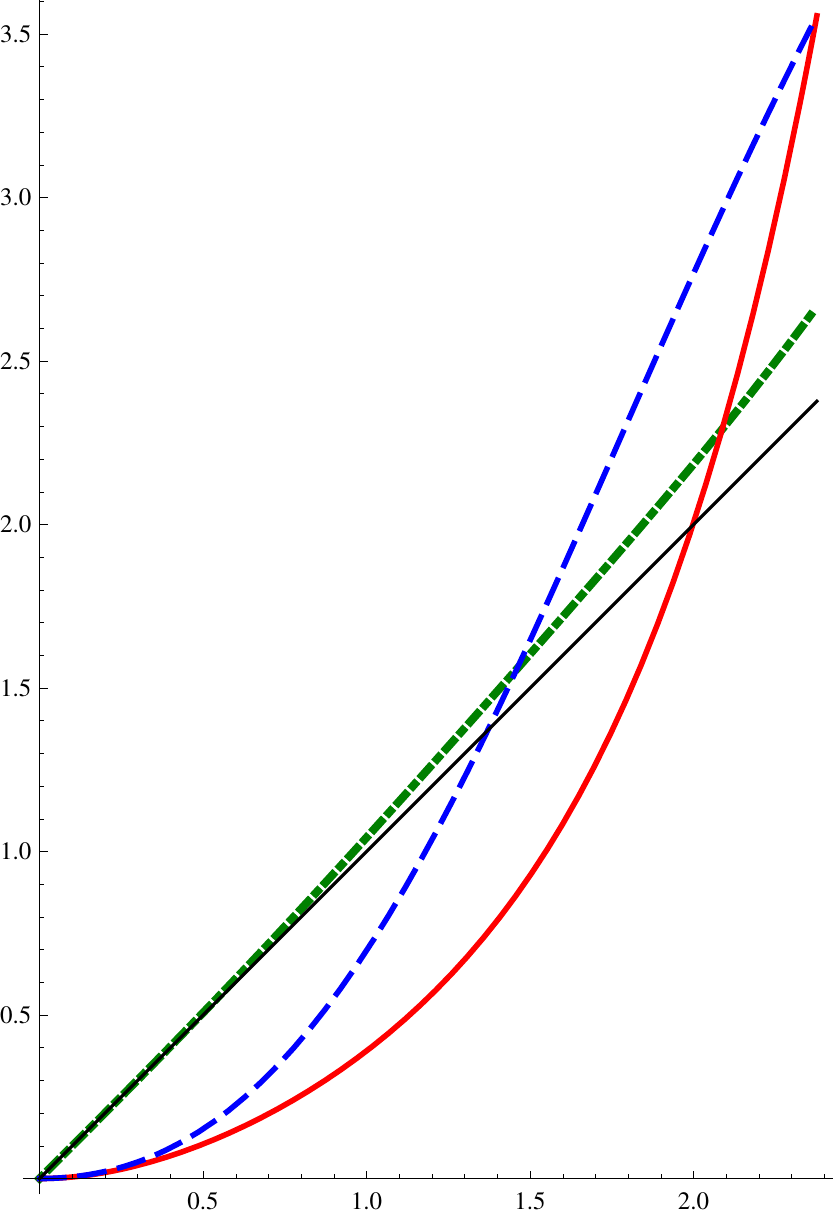}
\parbox{12.5cm}{\caption{Kullback-Leibler divergences compared to the Fisher distance along the geodesics of Figure~\ref{fig:circFisher} connecting the PDF's $A$ to $F$ (left) and $A$ to $B$ (right).}\label{fig:KB}}
\end{center}
\end{figure}

Figure \ref{fig:KB} (left) shows the graphics of the mappings  $g(d)=KL(A\| Y)$ (red continuous curve),  $g(-d)=KL(Y\| A)$ (blue dashed curve),
and the symmetrized $d_{KL}(A, Y)$ (green dot-dashed curve) when $Y$ goes  from $A$ to $F$ in Figure~\ref{fig:circFisher}, compared to the Fisher distance $d$ (identity), which varies in the interval $[0,2.3769]$. 

It is straightforward in this case to prove that the symmetrized Kullback-Leibler approaches the Fisher distance for small $d$. In fact, this result is more general, it also holds for multivariate normal distributions when $P$ approaches $Q$ in the parameter space \cite{CRH}.

Figure \ref{fig:KB} (right) displays the graphics of the mappings $KL(A\| Y)$ (red continuous curve),  $KL(Y\| A)$ (blue dashed curve),
and the symmetrized $d_{KL}(A, Y)$ (green dot-dashed curve), compared to the Fisher distance  $d$ (identity) varying  in the interval $[0,2.3769]$, with $Y$ going from $A$ to $B$ along the geodesic path of Figure~\ref{fig:circFisher}. 

\section{Fisher information geometry of multivariate normal distributions}
\label{sec:fishermult}

For more general $p$-variate PDF's, defined by an $n$-dimensional parameter space, 
the coefficients of the Fisher  matrix are given by
\[
g_{ij}(\negr{\beta})=\int\limits_{\R^p}f(\negr{x},\negr{\beta})\frac{\partial\ln
f(\negr{x},\negr{\beta})}{\partial\beta_{i}}\frac{\partial\ln
f(\negr{x},\negr{\beta})}{\partial \beta_{j}}d\negr{x}.
\]

The previous analysis can be extended to independent $p$-variate normal distributions:
\[
f(\negr{x},\negr{\mu}, \Sigma)=(2\pi)^{\frac{-p}{2}}(\det\Sigma)^{\frac{-1}{2}}%
\exp\left(\frac{-1}{2}(\negr{x}-\negr{\mu})^T\Sigma^{-1}(\negr{x}-\negr{\mu})\right),
\]
where%
\begin{align*}
\negr{x}&  = (x_{1},x_{2},\dots ,x_{p})^{T},\\ \negr{\mu} & = (\mu_{1},\mu_{2},\dots ,\mu_{p})^{T}\text{ (mean vector) and}\\
\Sigma &   \mbox{ is the covariance matrix (symmetric positive definite $p \times p$ matrix).}
\end{align*}

Note that, for general multivariate normal distributions, the parameter space has dimension $n=p+ p(p+1)/2$. 

\subsection{Round Gaussian distributions}\label{subsec:round}

If  $\Sigma  =\sigma^2 I$ (scalar covariance matrix), 
the set of all such distributions can be identified with the half $(p+1)$-dimensional space, $\Hh^{p+1}_F$, parametrized by
$\beta=(\mu_{1},\mu_{2},\ldots,\mu_{p},\sigma)$ and the
Fisher information matrix is:%

\[
\left[  g_{ij}\right]_F = \left[
\begin{array}
[c]{ccccc}%
\frac{1}{\sigma^{2}} & 0 & & & \\ 0 & \frac{1}{\sigma^{2}} & & & \\
& & \ddots & & \\ & & & \frac{1}{\sigma^2} & 0
\\
& & & 0 & \frac{2p}{\sigma^{2}}%
\end{array}
\right].
\]

We have again similarity with the matrix of the Poincar\'e model metric in the $(p+1)$-dimensional half space $\Hh^{p+1}$,
\[
\left[  g_{ij}\right]_H  = \left[
\begin{array}
[c]{ccccc}%
\frac{1}{\sigma^{2}} & 0 & & & \\ 0 & \frac{1}{\sigma^{2}} & & & \\
& & \ddots & & \\ & & & \frac{1}{\sigma^2} & 0
\\
& & & 0 & \frac{1}{\sigma^{2}}%
\end{array}
\right],
\]
and the similarity transformation 
\[
\Psi: {\Hh^{p+1}_F} \longrightarrow \Hh^{p+1}, \Psi(\mu_{1},\mu_{2},\ldots,\mu_{p},\sigma) =(\mu_{1}/\sqrt{2p},\mu_{2}/\sqrt{2p},\dots,\mu_{p}/\sqrt{2p},\sigma). 
\]

For $\boldsymbol \mu_1 =(\mu_{11},\mu_{12},\dots,\mu_{1p})$ and $\boldsymbol \mu_2 =(\mu_{21},\mu_{22},\dots,\mu_{2p})$ we have a closed form for the Fisher distance between the respective Gaussian PDF's:
\begin{equation}
\begin{array}{l}
d_{F,r}((\boldsymbol{\mu}_{1},\sigma_{1}),(\boldsymbol{\mu}_{2},\sigma_{2}))=\sqrt{2p} \,d_{H}\left(\left(\frac{\boldsymbol{\mu}_{1}%
}{\sqrt{2p}},\sigma_{1}\right),\left(\frac{\boldsymbol{\mu}_{2}}{\sqrt{2p}},\sigma_{2}\right)\right)\\
  = \sqrt{2p}\ln\dfrac{\left\vert
(\frac{\boldsymbol \mu_{1}}{\sqrt{2p}},\sigma_{1})-(\frac{\boldsymbol \mu_{2}}{\sqrt{2p}},-\sigma
_{2})\right\vert +\left\vert
(\frac{\boldsymbol \mu_{1}}{\sqrt{2p}},\sigma_{1})-(\frac
{\boldsymbol \mu_{2}}{\sqrt{2p}},\sigma_{2})\right\vert }{\left\vert (\frac{\boldsymbol \mu_{1}}%
{\sqrt{2p}},\sigma_{1})-(\frac{\boldsymbol \mu_{2}}{\sqrt{2p}},-\sigma_{2})\right\vert
-\left\vert (\frac{\boldsymbol \mu_{1}}{\sqrt{2p}},\sigma_{1})-(\frac{\boldsymbol \mu_{2}}{\sqrt{2p}%
},\sigma_{2})\right\vert }\end{array}
\label{eq:distround}
\end{equation}
where $| \cdot |$ is the standard Euclidean vector norm and the subindex $r$ stands for round distributions.

The geodesics in the parameter space $(\boldsymbol \mu, \sigma)$ between two round $p$-variate  Gaussian distributions are contained in planes orthogonal to the hyperplane $\sigma=0$, and are
either a line ($\mathbf \mu$ = constant) or a  half ellipse with eccentricity $\sqrt{2}$, centered at this hyperplane.

\subsection{Diagonal Gaussian distributions}\label{subsec:diag}

For general $\Sigma  =\mbox{diag}\text{ }(\sigma_{1}^{2},\sigma_{2}^{2},\dots ,\sigma_{p}^{2})\text{ (diagonal covariance matrix)}, \sigma_{i}>0, \forall i$, 
the set of all independent multivariate normal distributions is
parametrized by an intersection of half-spaces in $\mathbb{R}^{2p}$
($\beta=(\mu_{1},\sigma_{1},\mu_{2},\sigma_{2},\dots ,\mu_{p},\sigma_{p}),\sigma_{i}>0$) so the
Fisher information matrix is:%
\[
\left[  g_{ij}\right]_F  = \left[
\begin{array}
[c]{ccccc}%
\frac{1}{\sigma_{1}^{2}} & 0 & \cdots & 0 & 0\\ 0 &
\frac{2}{\sigma_{1}^{2}} & \cdots & 0 & 0 \\ \vdots & \vdots & \ddots & \vdots &
\vdots\\ 0 & 0 & \cdots & \frac{1}{\sigma_{p}^{2}} & 0\\
0 & 0 & \cdots & 0 & \frac{2}{\sigma_{p}^{2}}%
\end{array}
\right].
\]
We can show that, in this case, the metric is a product metric on the space ${\Hh^{2p}_F}$ and therefore we have the following closed form for the Fisher distance between the respective Gaussian PDFs:
\begin{multline}
d_{F,d}((\mu_{11},\sigma_{11},\dots,\mu_{1p},\sigma_{1p}),((\mu_{21},\sigma_{21},\dots,\mu_{2p},\sigma_{2p}))= \\ \sqrt{2}d_{{\Hh^{2}}^p}\left(\left(\dfrac{\mu_{11}}{\sqrt{2}},\sigma_{11},\dots,\dfrac{\mu_{1p}}{\sqrt{2}},\sigma_{1p}\right),\left(\dfrac{\mu_{21}}{\sqrt{2}},\sigma_{21},\dots,\dfrac{\mu_{2p}}{\sqrt{2}},\sigma_{2p}\right)\right),
\end{multline}
that is,
\begin{multline}
d_{F,d}((\mu_{11},\sigma_{11},\dots,\mu_{1p},\sigma_{1p}),((\mu_{21},\sigma_{21},\dots,\mu_{2p},\sigma_{2p})) = \\ 
= \sqrt{\sum_{i=1}^p 2 d_{\Hh^2}\left(\left(\frac{\mu_{1i}}{\sqrt{2}},\sigma_{1i}\right),\left(\frac{\mu_{2i}}{\sqrt{2}},\sigma_{2i}\right)\right)^2}  \\
=\sqrt{2\sum_{i=1}^p \left(\ln\dfrac{\left\vert (\frac{\mu_{1i}}{\sqrt{2}},\sigma_{1i})-(\frac{\mu_{2i}}{\sqrt{2}},-\sigma_{2i})\right\vert +\left\vert (\frac{\mu_{1i}}{\sqrt{2}},\sigma_{1i})-(\frac{\mu_{2i}}{\sqrt{2}},\sigma_{2i})\right\vert }{\left\vert (\frac{\mu_{1i}}{\sqrt{2}},\sigma_{1i})-(\frac{\mu_{2i}}{\sqrt{2}},-\sigma_{2i})\right\vert -\left\vert (\frac{\mu_{1i}}{\sqrt{2}},\sigma_{1i})-(\frac{\mu_{2i}}{\sqrt{2}},\sigma_{2i})\right\vert} \right)^2}%
\label{eq:distdiag}
\end{multline}
where $| \cdot |$ is the standard Euclidean vector norm and the subindex $d$ stands for diagonal distributions.

These matrices induce a metric of constant negative mean
curvature (i.e. a hyperbolic metric) which is equal to
$\frac{-1}{p(p+1)}$ in case of round distributions (\S\ref{subsec:round}) and to
$\frac{-1}{2(2p-1)}$ in case of diagonal distributions (\S\ref{subsec:diag}). Expressions for the distance and
other geometric properties can be deduced using results on product
of Riemannian manifolds and relations with Poincar\'{e} models
for hyperbolic spaces.

\subsection{General Gaussian distributions}
\label{subsec:general}

\begin{figure}[ht]
\begin{center}
\parbox{5cm}{
\includegraphics[width=5cm]{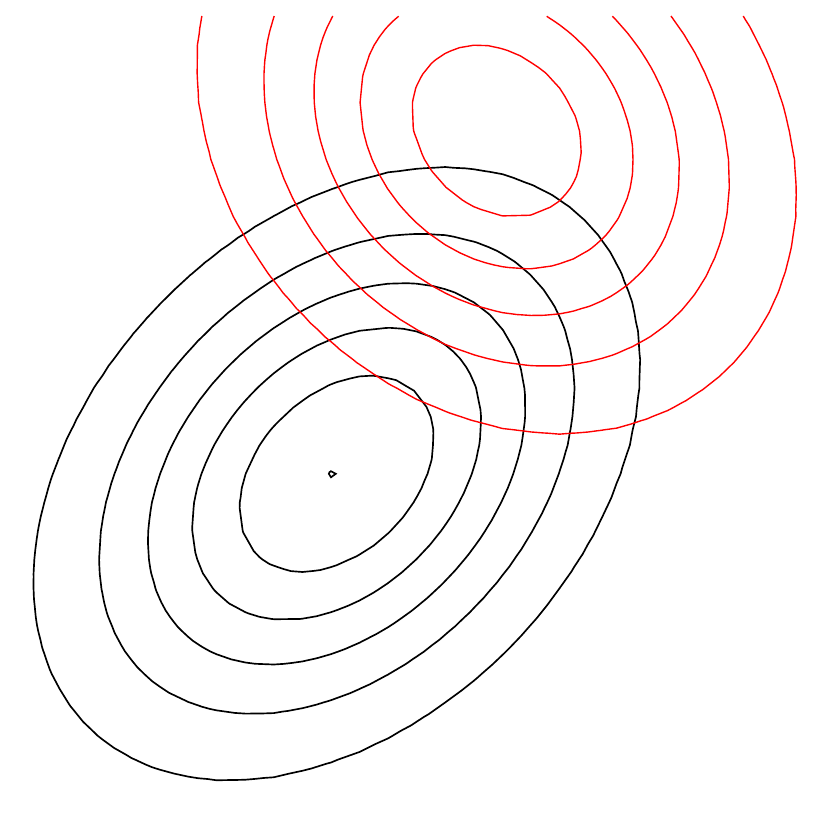} }
\parbox{8cm}{
\includegraphics[width=8cm]{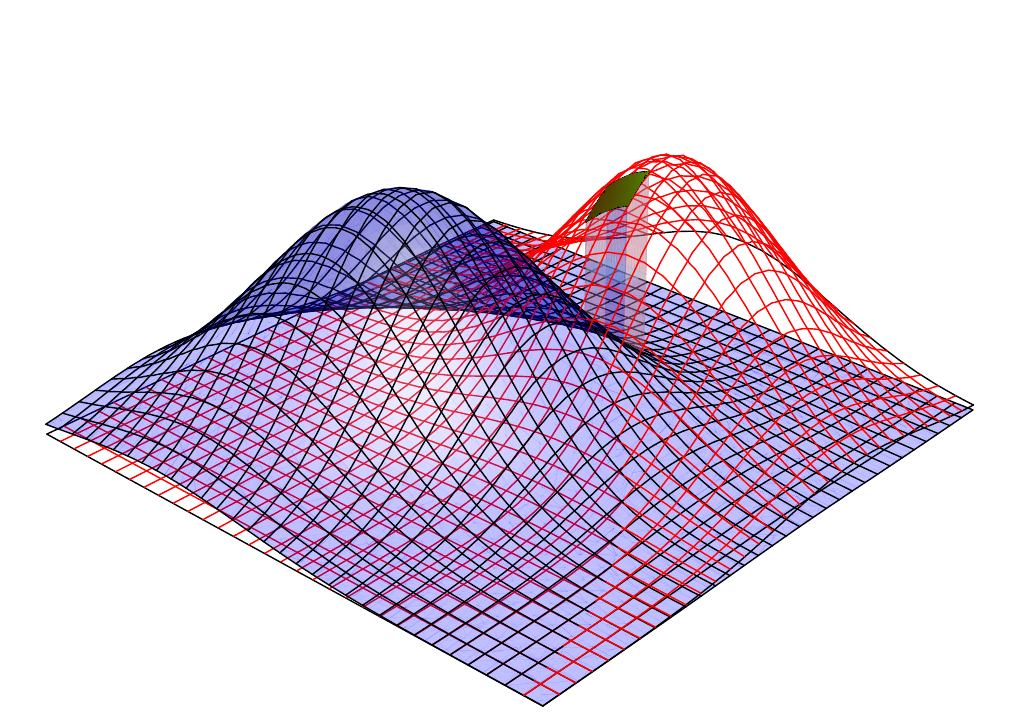} }
\parbox{12.5cm}{\caption{Bivariate normal distributions: level sets~(left) 
and representation in the upper half-space~(right).}\label{fig:normalBiv}}
\end{center}
\end{figure}

For general $p$-variate normal distributions (given by any symmetric
positive definite covariance matrices) the analysis is much more
complex as pointed out in \cite{atkinson} and far from being fully
developed. From the Riemannian geometry viewpoint this is due to the
fact that not all the sectional curvatures of their natural
parameter space (which is a $(p + p(p+1)/2)$-dimensional manifold) provided with the Fisher metric are constant. As an example, for $p=2$ we may
parametrize the general (elliptical) 2-variate normal distributions by $\beta=(\sigma_{1},\sigma_{2},\mu_{1} ,\mu_{2},u)$ \ where $\sigma_{1}^2,\sigma_{2}^2$
are the eigenvalues and $u$ the turning angle of the eigenvectors
of $\Sigma$. The level sets of a pair of such PDF's are families of rotated ellipses, see Figure~\ref{fig:normalBiv}. 

The Fisher matrix which induces the distance in this parameter space can be deduced as%
{\small\[
\left[  g_{ij}\right]_F = \left[\begin{array}
[c]{ccccc}%
\frac{2}{\sigma_{1}^{2}} & 0 & 0 & 0 & 0\\ 0 &
\frac{2}{\sigma_{2}^{2}} & 0 & 0 & 0\\ 0 & 0 &
\frac{\cos^{2}(u)}{\sigma_{1}^{2}}+\frac{\sin^{2}(u)}{\sigma_{2}^{2}}
&
\frac{\sin(2u)}{2}(\frac{1}{\sigma_{1}^{2}}+\frac{1}{\sigma_{2}^{2}})
& 0\\ 0 & 0 &
\frac{\sin(2u)}{2}(\frac{1}{\sigma_{1}^{2}}+\frac{1}{\sigma_{2}^{2}})
&
\frac{\cos^{2}(u)}{\sigma_{1}^{2}}+\frac{\sin^{2}(u)}{\sigma_{2}^{2}}
& 0\\
0 & 0 & 0 & 0 & \frac{(\sigma_{1}^{2}-\sigma_{2}^{2})^2
}{\sigma_{2}^{2}\sigma_{1}^{2}}
\end{array}\right].
\]}
We could not derive a general closed form for the associated Fisher distance in this parameter space.
Here, like in most multivariate cases, numerical approaches must be used to estimate the Fisher distance. In these approaches, the symmetrized Kullback-Leibler can be used to estimate the Fisher distance between nearby points in the parameter space~\cite{CRH}. Lower and upper bounds for the Fisher distance can also be found through an isometric embedding of the multivariate Gaussian distribution space into the Riemannian manifold of the positive definite symmetric matrices with the Siegel metric \cite{calvo90,calvo91}.

A special instance of the  bivariate model desribed above is given by the set of points with fixed means $\mu_1$, $\mu_2$ and turning angle $u=0$. Using the characterization of geodesics as solutions of a second order differential equation \cite{Manfredo}, we can assert that this two-dimensional submanifold is totally geodesic
(i.e. all the geodesics between two of such points are contained in this submanifold). Therefore,
the Fisher distance can be calculated as in (\ref{eq:distdiag}):
\begin{equation}
d_F((\sigma_{11},\sigma_{12},\mu_{1} ,\mu_{2},0),(\sigma_{21},\sigma_{22},\mu_{1} ,\mu_{2},0))=\sqrt{2}
\sqrt{  \left(\ln \left(\frac{\sigma_{11}}{\sigma_{12}} \right) \right)^2 +
  \left(\ln \left(\frac{\sigma_{21}}{\sigma_{22}} \right) \right)^2}.
\label{eq:distgeneral}
\end{equation}

If we consider the $(p(p+1)/2$-dimensional statistical model of $p$-variate normal PDF's with fixed mean $\negr{\mu}$ and general covariance matrix $\Sigma$, the induced Fisher distance can be deduced \cite{atkinson} as
\begin{equation}
d_F^2((\negr{\mu}, \Sigma_1),(\negr{\mu}, \Sigma_2)) = \frac{1}{2} \sum_{j=1}^p (\ln \lambda_j)^2,
\label{eq:autovalores}
\end{equation}
where $\lambda_j$ are the eigenvalues of matrix $ (\Sigma_1)^{-1}\Sigma_2$ (i.e. $\lambda_j$  are the roots of the equation \linebreak $\det( (\Sigma_1)^{-1}\Sigma_2 - \lambda I) = 0$). 
Note that, for $p=1$,  the expression (\ref{eq:autovalores}) reduces to (\ref{eq:verticaldistance}).

Moreover, by restricting  (\ref{eq:autovalores})  to the set of distributions with diagonal covariance matrices, the induced metric is the same as the metric restricted to distributions with fixed mean $\negr{\mu}$ (cf.~\S\ref{subsec:diag}).

\section{Final remarks}
\label{sec:final}

We have presented  a geometrical view of the Fisher distance,
 focusing on the parameters that describe the normal distributions, to widen the range of possible interpretations for the prospective applications of information geometry.

By exploring the  two dimensional statistical model of the Gaussian (normal) univariate PDF, we have
 employed  results from the classical hyperbolic geometry to
 derive closed forms for the Fisher distance in the most commonly used parameters. A relationship with the Kullback-Leibler measure of divergence was derived as well. The multivariate normal PDF's were also analyzed from the geometrical standpoint and closed forms for the Fisher distance were derived in special instances.


\end{document}